\documentclass[3p,twocolumn]{elsarticle}

\usepackage{amsmath}  
\usepackage{amsfonts}
\usepackage{amssymb}
\usepackage{natbib} 
\usepackage{setspace}
\usepackage{graphicx} 
\usepackage{xspace}
\usepackage{cancel}
\usepackage{color}
\usepackage{float}
\usepackage{url}
\usepackage{soul}

\usepackage{ulem}

\newcommand{\mi}{\mathrm{i}} 

\newcommand{\dx}[1]{\hspace{-0.4em}\ensuremath{\mathrm{d}#1}\,}

\newcommand{\eqn}[1]{Eq.~(\ref{#1})}
\newcommand{\fig}[1]{Fig.~\ref{#1}}
\newcommand{\tab}[1]{Table~\ref{#1}}

\def\Kbar{\bar{K}}
\def\K0bar{\overline{K}^{0}}
\def\Dbar{\bar{D}}
\def\D0bar{\bar{D}^{0}}
\def\Dsta{D^{*}}

\def\etac{\eta_{c}}

\newcommand{\be}{\begin{equation}}
\newcommand{\ee}{\end{equation}}
\newcommand{\bge}{\begin{equation}}
\newcommand{\ene}{\end{equation}}
\newcommand{\bea}{\begin{eqnarray}}
\newcommand{\eea}{\end{eqnarray}}
\newcommand{\bg}{\begin{eqnarray}}
\newcommand{\en}{\end{eqnarray}}

\begin{document}

\title{
  \vspace{-25mm}
  \begin{flushright}
    LFTC-20-4/56, ADP-20-15/T1125
  \end{flushright}
  \vspace{15mm}
  $\eta_{c}$-nucleus bound states}


\author[add1,add2]{J.J.~Cobos-Mart\'{\i}nez\corref{corr}}
\ead{jesus.cobos@fisica.uson.mx}
\author[add3]{K.~Tsushima}
\ead{kazuo.tsushima@gmail.com}
\author[add4]{G.~Krein}
\ead{gkrein@ift.unesp.br}
\author[add5]{A.W.~Thomas}
\ead{anthony.thomas@adelaide.edu.au}

\address[add1]{Departamento de F\'isica, Universidad de Sonora, Boulevard
Luis Encinas J. y Rosales, Colonia Centro, Hermosillo, Sonora 83000, M\'exico}
\address[add2]{C\'atedra CONACyT, Departamento de F\'{\i}sica, Centro de Investigaci\'on
  y de Estudios Avanzados del Instituto Polit\'ecnico Nacional, Apartado Postal 14-740,
  07000, Ciudad de M\'exico, M\'exico}
\address[add3]{Laborat\'orio de F\'{\i}sica Te\'orica e Computacional-LFTC, 
Universidade Cruzeiro do Sul and Universidade Cidade de S\~ao Paulo (UNICID),
01506-000, S\~ao Paulo, SP, Brazil}
\address[add4]{Instituto de F\'{\i}sica Te\'orica, Universidade Estadual 
Paulista, Rua Dr. Bento Teobaldo Ferraz, 271 - Bloco II, 01140-070, 
S\~ao Paulo, SP, Brazil}
\address[add5]{CSSM, School of Physical Sciences, University of Adelaide,
Adelaide SA 5005, Australia}

\cortext[corr]{Corresponding author}

\date{\today}

\begin{abstract}
$\etac$-nucleus bound state energies are calculated for various nuclei. 
Essential input for the calculations, namely the medium-modified $D$ and $\Dsta$ meson masses, as well as the density distributions in nuclei, are calculated 
within the quark-meson coupling (QMC) model. 
The attractive potentials for the $\etac$ meson in the nuclear medium originate from the in-medium enhanced $D\Dsta$ loops in the $\etac$ self-energy.
Our results suggest that the $\etac$ meson should form bound states with all the
nuclei considered.
\end{abstract}


\maketitle

\date{\today}

\section{Introduction}

The study of the interactions of charmonium states, such as $\eta_c$ and $J/\Psi$,  
with atomic nuclei offers opportunities to 
gain new insight into the properties of the strong force and strongly 
interacting matter~\cite{Brodsky:1989jd,Wasson:1991fb}. Because charmonia and 
nucleons do not share light quarks, the Zweig rule suppresses interactions mediated by
the exchange of mesons made of light quarks. It is therefore vital to explore
other potential sources of attraction which could potentially lead  to binding of 
charmonia to atomic nuclei.

A large body of work looking for alternatives to the meson-exchange 
paradigm has accumulated over the last three decades~\cite{Hosaka:2016ypm,
Krein:2016fqh,Metag:2017yuh,Krein:2017usp}. There are works based on the 
charmonium color polarizability~\cite{Peskin:1979va,Kharzeev:1995ij}, 
responsible for long-range van der Waals type of forces~\cite{Kaidalov:1992hd,
Luke:1992tm,deTeramond:1997ny,Brodsky:1997gh,Ko:2000jx,Sibirtsev:2005ex,
Voloshin:2007dx,TarrusCastella:2018php}. Others employ charmed meson loops, 
with light quarks created from the vacuum~\cite{Ko:2000jx,Krein:2010vp,
Tsushima:2011kh,Tsushima:2011fg,Krein:2013rha}. There are studies based 
on QCD sum rules~\cite{Klingl:1998sr,Hayashigaki:1998ey,Kim:2000kj,Kumar:2010hs} 
and phenomenological potentials~\cite{Belyaev:2006vn,Yokota:2013sfa}. 
More recently, there appeared lattice QCD simulations of the binding of charmonia
to nuclear matter and finite nuclei~\cite{Beane:2014sda}, as well as light mesons and
baryons~\cite{Alberti:2016dru}. 
The lattice QCD simulations of Ref.~\cite{Beane:2014sda} have demonstrated that
quarkonium-nucleus bound states exist for $A<5$. Ref.~\cite{Beane:2014sda} also
infers a charmonium-nuclear matter binding energy $B^{NM}\sim 60$ MeV. 
However, these simulations have been performed at the flavor $SU(3)$-symmetric
point, with unphysical pion masses, $m_{\pi}\sim 805$~MeV.

Model studies have suffered from scarce experimental data on the low-energy 
charmonium-nucleon interaction. However, this situation started to change 
for the $J/\Psi$ case with the 
recent measurement, by the JLab GlueX Collaboration~\cite{Ali:2019lzf}, of the 
$\gamma \, p \rightarrow J/\Psi \, p$ total cross section near threshold.  
It will further improve with the completion of other close-to-threshold $J/\Psi$ 
photoproduction experiments at JLab~\cite{Stepanyan,Meziani:2016lhg}. 
Regarding production on nuclei, there is a JLab proposal to measure $J/\Psi$ 
photoproduction off the deuteron~\cite{Ilieva}. 
On the other hand, unfortunately, there are not
many experiments specially directed towards the $\eta_c$ meson and its binding to
nuclei, perhaps because it is more difficult to produce and detect. 
Recent studies on $\eta_c$ production in heavy ion collisions
($pp,\,pA,\,AA$) at the LHC have been carried out in Refs.~\cite{Aaij:2019gsn,Tichouk:2020dut,Tichouk:2020zhh,Goncalves:2018yxc,Klein:2018ypk} towards the experimental 
study of its underlying production mechanisms.
However, $J/\Psi$ measurements are a first step towards to experimental study
of binding of charmonia to nuclei. 
From the theory side, lattice QCD simulations of the free-space charmonium-nucleon interaction have become available
within the last decade~\cite{Yokokawa:2006td,Liu:2008rza,Kawanai:2010ev,Kawanai:2010ru,Skerbis:2018lew}. 
Unfortunately they have either been quenched or used large pion masses, which
therefore require extrapolation to the physical mass~\cite{TarrusCastella:2018php}. 

Although crucial for constraining models, experimental knowledge of the free-space
charmonium-nucleon interaction is not enough for assessing the likelihood of
charmonium binding in nuclei. The overwhelming evidence that the internal
structure of hadrons changes in  medium~\cite{Krein:2016fqh,Krein:2017usp,Hayano:2008vn,Leupold:2009kz} must be taken
into account when addressing charmonium in nuclei. As shown in previous 
studies~\cite{Ko:2000jx,Krein:2010vp,Tsushima:2011kh,Tsushima:2011fg,Krein:2013rha},
the effect of the nuclear mean fields on subthreshold $D\Dbar$ states is of particular 
relevance. Those studies have revealed that modifications induced by the strong nuclear mean 
fields on the $D$ mesons' light-quark content enhance the self-energy in such a way 
as to provide an attractive $J/\Psi-$nucleus 
effective potential. In the present paper we extend our previous study on the 
$J/\Psi$-nucleus bound states~\cite{Krein:2010vp,Tsushima:2011fg} to the case of $\eta_c$ 
charmonium. $\etac$-nucleus bound states have also been predicted in other
approaches~\cite{Brodsky:1989jd,Wasson:1991fb,deTeramond:1997ny,Klingl:1998sr,Kumar:2010hs}, 
albeit with predictions for the binding energies varying over a wide range. 

It is worth stressing that compared to the situation for the lighter $\phi$ 
meson~\cite{Ko:1992tp,Klingl:1997tm,Oset:2000eg,Cabrera:2002hc,Gubler:2015yna,
Cobos-Martinez:2017vtr,Cobos-Martinez:2017woo,Cobos-Martinez:2017onm,Cobos-Martinez:2017fch,
Cobos-Martinez:2019kln}, which couples strongly to above-threshold $K\Kbar$ states, 
the charmonium states are expected to have
a small width in medium and therefore the signal for the formation of such bound states may be experimentally cleaner.

This paper is organized as follows. In Sec.~\ref{sec:etacnm} we
discuss the computation and present results for the mass shift 
of the $\eta_{c}$ in symmetric nuclear matter.
Using the results of Sec.~\ref{sec:etacnm}, together with the
density profiles of the nuclei calculated within
the quark-meson coupling model,  in Sec.~\ref{sec:etacbs} 
we present results for the scalar $\eta_{c}$-nucleus potentials, 
as well as the corresponding bound state energies.
Finally, Sec.~\ref{sec:summary} is devoted to a summary and
conclusions.

\section{\label{sec:etacnm} 
Calculation of the $\etac$ scalar potential in symmetric nuclear matter}

For the computation of the $\etac$ scalar potential in nuclear matter we use an
effective Lagrangian approach at the hadronic level~\cite{Klingl:1996by}, which is an $SU(4)$-flavor extension of light-flavor chiral-symmetric 
Lagrangians of pseudoscalar and vector mesons~\cite{Lin:1999ve}.
  
The extracted interaction Lagrangian density for the $\etac D D^{*}$ vertex
is given by 
\bea
\label{eqn:LetaDDsta}
&&\hspace{-10ex}\mathcal{L}_{\etac D D^{*}} = 
\nonumber\\
&&\hspace{-6ex}\mi g_{\etac D D^{*}}(\partial_{\mu} \etac)
\left[\Dbar^{*\mu}\cdot D - \Dbar\cdot D^{*\mu}\right]\, 
\nonumber\\
&&\hspace{-6ex}-\mi g_{\etac D D^{*}}(\etac)
\left[\Dbar^{*\mu}\cdot (\partial_\mu D) - (\partial_\mu \Dbar)\cdot D^{*\mu}\right]\, , 
\eea
where $D^{(*)}$ represents the $D^{(*)}$-meson field isospin doublet,
and $g_{\etac D D^{*}}$ is the coupling constant to be specified below.

We employ the effective interaction Lagrangian Eq.~(\ref{eqn:LetaDDsta}) to
compute the $\etac$ and self energy in vacuum and symmetric nuclear matter, following our previous works~\cite{Krein:2010vp, Tsushima:2011kh,
Tsushima:2011fg, Krein:2013rha,Cobos-Martinez:2017vtr,
Cobos-Martinez:2017woo, Cobos-Martinez:2017onm,Cobos-Martinez:2017fch,
Cobos-Martinez:2019kln}, and considering only they would be dominant
$DD^{*}$ loop. The $\eta_c$ self-energy is thus given by 
\begin{equation}
    \label{eqn:etac_se}
    \Sigma_{\etac}(k^{2})= \frac{8g_{\etac D D^{*}}^{2}}{\pi^{2}}\int_{0}^{\infty}
    \dx{k}k^{2}I(k^{2})
\end{equation}
for an $\etac$ at rest, where
\begin{align}
I(k^{2})&= \left. \frac{m_{\etac}^{2}(-1+k^{0\,2}/m_{D^{*}}^{2})}
{(k^{0}+\omega_{D^{*}})(k^{0}-\omega_{D^{*}}) 
(k^{0}-m_{\etac}-\omega_{D})}\right|_{k^{0}=m_{\etac}-\omega_{D^{*}}} 
\nonumber \\
+&\left. \frac{m_{\etac}^{2}(-1+k^{0\,2}/m_{D^{*}}^{2})}
{(k^{0}-\omega_{D^{*}})(k^{0}-m_{\etac}+\omega_{D}) 
(k^{0}-m_{\etac}-\omega_{D})}\right|_{k^{0}=-\omega_{D^{*}}},  
\end{align}
and $\omega_{D^{(*)}}=(k^{2}+m_{D^{(*)}}^{2})^{1/2}$, with
$k=|\vec{k}|$.
The integral in \eqn{eqn:etac_se} is divergent and thus needs 
regularization. For this purpose we employ a phenomenological
vertex form factor
\begin{equation}
    \label{eqn:FF}
    u_{D^{(*)}}(k^{2})=
    \left(\frac{\Lambda_{D^{(*)}}^{2} + m_{\etac}^{2}}
    {\Lambda_{D^{(*)}}^{2}+4\omega_{D^{(*)}}^{2}(k^{2})}
    \right)^{2},
\end{equation}
with cutoff parameter $\Lambda_{D^{(*)}}$, as in
Refs.~\cite{Krein:2010vp, Tsushima:2011kh, Tsushima:2011fg, Krein:2013rha,
  Cobos-Martinez:2017vtr, Cobos-Martinez:2017woo, Cobos-Martinez:2017onm,
  Cobos-Martinez:2017fch, Cobos-Martinez:2019kln}. 
Thus, to regularize \eqn{eqn:etac_se} we will introduce the
factor $ u_{D}(k^{2})u_{D^{*}}(k^{2})$ into the integrand.

The cutoff parameter $\Lambda_D$ (we use $\Lambda_{D^{*}}=\Lambda_{D}$) is an 
unknown input to our calculation. However, it may be determined phenomenologically using, for 
example, a quark model. 
In fact, in Ref.~\cite{Krein:2010vp} its value has been estimated to be
$\Lambda\approx 2500\,\textrm{MeV}$, and it serves us as a reasonable guide to quantify
the sensitivity of our results to its value. Therefore we vary it
over the the interval 1500-3000 MeV.

Because $SU(4)$ flavor symmetry is strongly 
broken in Nature, we use experimental values for the meson 
masses~\cite{Tanabashi:2018oca} and empirically known values 
for the coupling constants, as explained below.
For the $D$ meson mass, we take the averaged masses of the neutral
and charged states, and similarly for the $D^{*}$. Thus
$m_{D}=1867.2$ MeV and $m_{D^{*}}=2008.6$ MeV.
For the coupling constants, $g_{\etac D D^{*}}=0.60\,g_{\psi D D}$ was 
obtained in Ref.~\cite{Lucha:2015dda} as the residue at the
poles of suitable form factors using a dispersion formulation of
the relativistic constituent quark model, where $g_{\psi D D}=7.64$ 
was estimated in Ref.~\cite{Lin:1999ad} using the vector meson 
dominance model and isospin symmetry.
In this study we use the coupling constant, 
$g_{\etac D D^{*}}=(0.60/\sqrt{2})\,g_{\psi D D} \simeq 0.424\,g_{\psi D D}$, 
where the factor ($1/\sqrt{2}$) is introduced to give   
a larger $SU(4)$ symmetry breaking effect than Ref.~\cite{Lucha:2015dda}.
In this connection
we mention that recent investigations of $SU(4)$ flavor symmetry
breaking in hadron couplings of charmed hadrons to light mesons
are not conclusive; while two studies based on Schwinger-Dyson
equations of QCD find large deviations from $SU(4)$
symmetry~\cite{ElBennich:2011py,El-Bennich:2016bno}, studies 
using QCD sum rules~\cite{Navarra:1998vi,Khodjamirian:2011jp}, 
a constituent quark model~\cite{Fontoura:2017ujf} and a
holographic QCD model~\cite{Ballon-Bayona:2017bwk} find moderate
deviations.

We are interested in the difference between the in-medium,
$m_{\etac}^{*}$, and vacuum, $m_{\etac}$, masses of the $\etac$,
\begin{equation}
    \Delta m_{\etac}= m_{\etac}^{*}-m_{\etac},
\end{equation}
with the masses obtained self-consistently from
\begin{equation}
    m_{\etac}^{2}= (m_{\etac}^{0})^{2} + \Sigma_{\etac}(k^{2}=m_{\etac}^{2}),
\end{equation}
where $m_{\etac}^{0}$ is the bare $\etac$ mass and
$\Sigma_{\etac}(k^{2})$ is given by \eqn{eqn:etac_se}. 
The $\Lambda_{D}$-dependent $\etac$-meson bare mass,  
$m_{\etac}^{0}$, is fixed by fitting the physical $\etac$-meson 
mass, $m_{\etac}=2983.9$ MeV.

The in-medium $\etac$ mass is obtained in a similar way, with the
self-energy calculated with the medium-modified $D$ and $D^{*}$
masses.
The nuclear density dependence of the $\etac$-meson mass is determined
by the intermediate-state $D$ and $D^{*}$ meson interactions with
the nuclear medium through their medium-modified masses.
The in-medium masses $m_{D}^{*}$ and $m_{D^{*}}^{*}$ are 
calculated within the quark-meson coupling (QMC)
model~\cite{Krein:2010vp,Tsushima:2011kh}, in which effective
scalar and vector meson mean fields couple to the light $u$
and $d$ quarks in the charmed mesons~\cite{Krein:2010vp,
Tsushima:2011kh}.
The QMC model has proven to be very successful in studying the
properties of hadrons in nuclear matter and finite 
nuclei~\cite{Saito:2005rv,Guichon:2006er,Stone:2016qmi,Tsushima:1997df,Guichon:1989tx}. 
This model considers infinitely large, uniformly symmetric,
spin-isospin-saturated nuclear matter in its rest frame, where all
the scalar and vector mean field potentials, which are responsible
for the nuclear many-body interactions, become constant in the 
Hartree approximation~\cite{Saito:2005rv,Tsushima:1997df,Guichon:1989tx}.

In \fig{fig:mDDsta} we present the resulting medium-modified masses
for the $D$ and $D^{*}$ mesons, calculated within the QMC model
~{\cite{Krein:2010vp}, as
a function of $\rho_{B}/\rho_{0}$, where $\rho_{B}$ is the baryon
density of nuclear matter and  $\rho_{0}=0.15$ fm$^{-3}$ is the 
saturation density of symmetric nuclear  matter.
The net reductions in the masses of the $D$ and $D^{*}$ mesons 
are nearly the same as a function of density, 
with each decreasing by around 60 MeV at $\rho_{0}$.

The behavior of the $D$ meson mass in medium (finite density and/or
temperature) has been studied in a variety of approaches.
Some of these~\cite{Hayashigaki:2000es,Azizi:2014bba,Wang:2015uya}
find a decreasing $D$ meson mass at finite baryon
density, while others~\cite{Suzuki:2015est,Park:2016xrw,Gubler:2020hft,Hilger:2008jg, Carames:2016qhr}, interestingly, find the opposite
behavior. 
However, it is important to note that none of the studies in nuclear
matter are constrained by the saturation properties of nuclear
matter, although it is constrained in the case of the present work.
Furthermore, some of these works employ a non relativistic approach 
and relativistic effects might be important.

\begin{figure}
\centering
\includegraphics[scale=0.275]{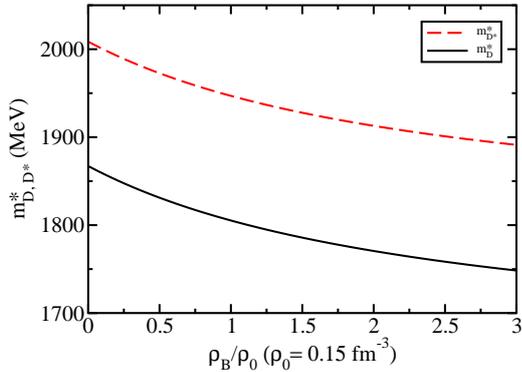}
\caption{\label{fig:mDDsta} In-medium $D$ and $D^{*}$ meson masses calculated within the QMC model. Adapted from Ref.~\cite{Krein:2010vp}}
\end{figure}
%

\begin{figure}
\centering
\includegraphics[scale=0.275]{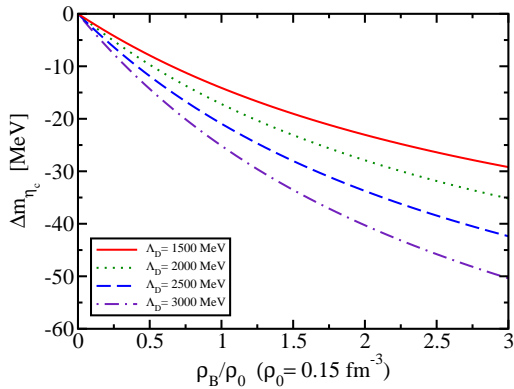}
\caption{\label{fig:Dmetac} $\etac$ mass shift as a function of
the nuclear matter density for various values of the cutoff parameter.}
\end{figure}
%
In \fig{fig:Dmetac}, we present the $\etac$-meson mass shift,
$\Delta m_{\eta_{c}}$, as a function of the nuclear matter density,
$\rho_{B}$, normalized to $\rho_{0}$, for four values of the cutoff
parameter $\Lambda_{D}$.
As can be seen from the figure, the effect of the in-medium $D$
and $D^{*}$ mass change is to shift the $\etac$ mass downwards.
This is because  the reduction in the $D$ and $D^{*}$ masses enhances
the $DD^{*}$-loop contribution in nuclear matter relative to that
in vacuum. This effect increases the larger the cutoff mass
$\Lambda_{D}$.

The results described above support a small downward mass shift for 
the $\etac$ in nuclear matter and open the possibility to study the
binding of $\etac$ mesons to nuclei, to which we turn our
attention in the next section.

\section{\label{sec:etacbs} $\etac$-nucleus bound states}

\begin{figure*}
  \begin{tabular}{c@{\hskip 3mm}c@{\hskip 3mm}c}
\includegraphics[scale=0.3]{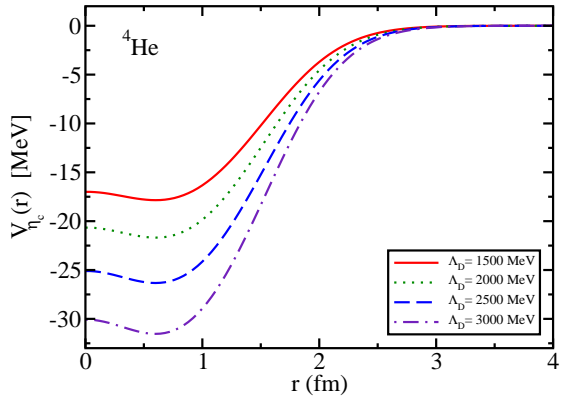} &  
\includegraphics[scale=0.3]{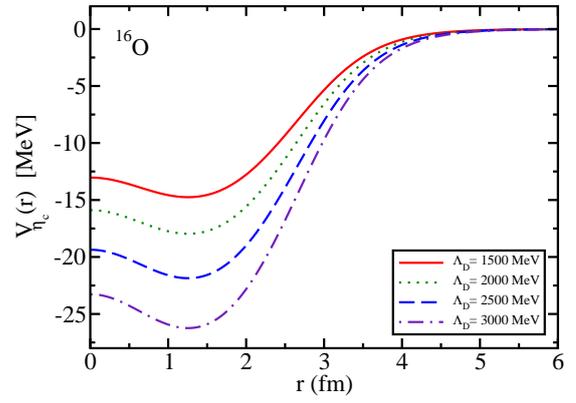} \\
\includegraphics[scale=0.3]{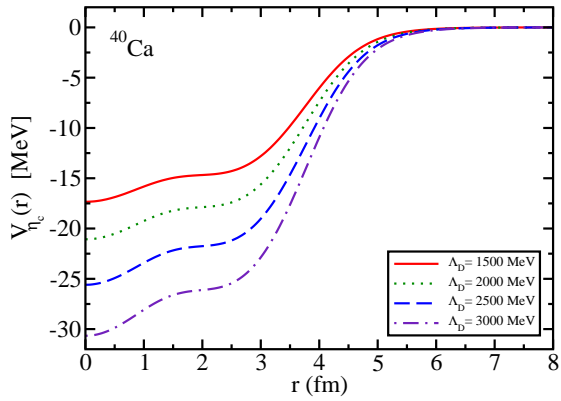} & 
\includegraphics[scale=0.3]{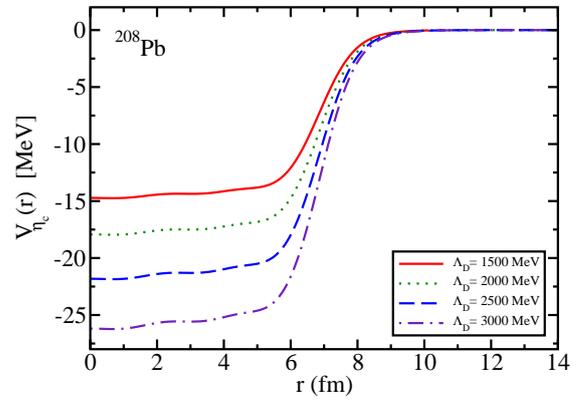} \\
  \end{tabular}
  \caption{\label{fig:VetacA}
    $\etac$-nucleus  potentials for various nuclei and values of the
    cutoff parameter $\Lambda_{D}$.}
\end{figure*}

We now discuss the situation where the $\etac$-meson is produced
inside a nucleus $A$ with baryon density distribution
$\rho_{B}^{A}(r)$.
The nuclei we consider here are $^{4}$He, $^{12}$C, $^{16}$O,
$^{40}$Ca, $^{48}$Ca, $^{90}$Zr, $^{197}$Au, and $^{208}$Pb. 
Their nuclear density distributions are also calculated within the 
QMC model, except for $^{4}$He, whose parametrization was 
obtained in Ref.~\cite{Saito:1997ae}.
Using a local density approximation, the $\etac$-meson potential 
within nucleus $A$ is given by 
\begin{equation}
\label{eqn:VetacA}
V_{\etac A}(r)= \Delta m_{\etac}(\rho_{B}^{A}(r)),
\end{equation}
\noindent where $r$ is the distance from the center of the nucleus.

In~\fig{fig:VetacA} we present the $\etac$-meson potentials
for a selection of the nuclei mentioned above and various values of the cutoff
parameter $\Lambda_{D}$.
From the figure one can see that the $\etac$ potential in nuclei is
attractive in all cases but its depth depends on the value of the 
cutoff parameter, being deeper the larger $\Lambda_{D}$ is. 
For example, it varies, from -18 MeV to -32 MeV for $^{4}$He and from
-15 MeV to -26 MeV for $^{208}$Pb, when the cutoff varies from 1500
MeV to 3000 MeV. This dependence is, indeed, an uncertainty in the
results obtained in our approach. 

%
\begin{table}[ht!]
\begin{center}
\scalebox{0.75}{
\begin{tabular}{ll|r|r|r|r}
  \hline \hline
  & & \multicolumn{4}{c}{Bound state energies} \\
  \hline
& $n\ell$ & $\Lambda_{D}=1500$ & $\Lambda_{D}=2000$ & $\Lambda_{D}=2500$ & 
$\Lambda_{D}= 3000$ \\
\hline
$^{4}_{\eta_{c}}\text{He}$
    & 1s & -1.49 & -3.11 & -5.49 & -8.55 \\
\hline
$^{12}_{\eta_{c}}\text{C}$
    & 1s & -5.91 & -8.27 & -11.28 & -14.79 \\
    & 1p & -0.28 & -1.63 & -3.69  & -6.33 \\
\hline
$^{16}_{\eta_{c}}\text{O}$
    & 1s & -7.35 & -9.92 & -13.15 & -16.87 \\
    & 1p & -1.94 & -3.87 & -6.48  & -9.63 \\
\hline
$^{40}_{\eta_{c}}\text{Ca}$
    & 1s & -11.26 & -14.42 & -18.31 & -22.73 \\
    & 1p & -7.19  & -10.02 & -13.59 & -17.70 \\
    & 1d & -2.82  &  -5.22 &  -8.36 & -12.09 \\
    & 2s & -2.36  &  -4.51 &  -7.44 & -10.98 \\
\hline
$^{48}_{\eta_{c}}\text{Ca}$
    & 1s & -11.37 & -14.46 & -18.26 & -22.58 \\
    & 1p & -7.83  & -10.68 & -14.23 & -18.32 \\
    & 1d & -3.88  & -6.40  & -9.63  & -13.41 \\
    & 2s & -3.15  & -5.47  & -8.54  & -12.17 \\
\hline
$^{90}_{\eta_{c}}\text{Zr}$
    & 1s & -12.26 & -15.35 & -19.14 & -23.43 \\
    & 1p & -9.88  & -12.86 & -16.53 & -20.70 \\
    & 1d & -7.05  & -9.87  & -13.38 & -17.40 \\
    & 2s & -6.14  & -8.87  & -12.29 & -16.24 \\
    & 1f & -3.90  & -6.50  & -9.81  & -13.65 \\
\hline
$^{197}_{\eta_{c}}\text{Au}$
      & 1s & -12.57 & -15.59 & -19.26 & -23.41 \\
      & 1p & -11.17 & -14.14 & -17.77 & -21.87 \\
      & 1d & -9.42  & -12.31 & -15.87 & -19.90 \\
      & 2s & -8.69  & -11.53 & -15.04 & -19.02 \\
      & 1f & -7.39  & -10.19 & -13.70 & -17.61 \\
\hline
$^{208}_{\eta_{c}}\text{Pb}$
    & 1s & -12.99 & -16.09 & -19.82 & -24.12 \\
    & 1p & -11.60 & -14.64 & -18.37 & -22.59 \\
    & 1d & -9.86  & -12.83 & -16.49 & -20.63 \\
    & 2s & -9.16  & -12.09 & -15.70 & -19.80 \\
    & 1f & -7.85  & -10.74 & -14.30 & -18.37 \\
\hline 
\hline
\end{tabular}
}
  \caption{\label{tab:etac-A-kg-be} $\etac$-nucleus bound
 state energies for different values of the cutoff parameter $\Lambda_{D}$.
 All dimensionful quantities are in MeV.}
\end{center}
\end{table}

Using the $\etac$-meson potentials obtained in this manner, we
next calculate the $\etac$-meson--nucleus bound state energies 
for the nuclei listed above by solving the Klein-Gordon equation
\begin{equation}
\label{eqn:kg}
\left(-\nabla^{2} + \mu^{2} + 2\mu V(\vec{r})\right)\phi_{\etac}(\vec{r})
= \mathcal{E}^{2}\phi_{\etac}(\vec{r}),
\end{equation}
\noindent where $\mu=m_{\etac}m_{A}/(m_{\etac}+m_{A})$ is the reduced
mass of the $\etac$-meson-nucleus system with $m_{\etac}$ $(m_{A}$)
the mass of the $\etac$-meson (nucleus $A$) in vacuum, and
$V(\vec{r})$ is the $\etac$-meson-nucleus potential given in 
\eqn{eqn:VetacA}. 

The bound state energies ($E$) of the $\etac$-nucleus system,
given by $E= \mathcal{E}-\mu$, where $\mathcal{E}$ is the energy
eigenvalue in \eqn{eqn:kg}, are calculated for four values of the
cutoff parameter $\Lambda_{D}$ and are listed in
\tab{tab:etac-A-kg-be}.
These results show that the $\etac$-meson is expected to form bound
states with all the nuclei studied and this prediction is independent of the value 
of the cutoff parameter $\Lambda_{D}$.
However, the particular values for the bound state energies are
clearly dependent on $\Lambda_{D}$, 
namely, each of them increases  in absolute value 
as $\Lambda_D$ increases. This was expected from the behavior of the
$\etac$ potentials, since these are deeper for larger values of the
cutoff parameter. Note also that the $\etac$ binds more strongly
to heavier nuclei.
We have also solved the Schr\"{o}edinger equation with the potential
\eqn{eqn:VetacA} to obtain the single-particle
energies~\cite{Krein:2017usp} and compared these
with those given in \tab{tab:etac-A-kg-be}. The results found in 
both cases are essentially the same.

Note that we have ignored the natural width of 32 MeV in free space of the  $\eta_c$ but 
understand this could be an issue related to the observability of the predicted bound states. 
Furthermore, we have no reason to believe the width will be suppressed in medium. Thus, even though 
it could be difficult to resolve the 
individual states, it should be possible to see that there
are bound states which is the main point of this work. It remains to be seen by how much the 
inclusion of a 
repulsive imaginary part will affect the predicted bound states. We believe this can be done in 
future work.

Another effect that could potentially have important consequences for 
the formation of the bound states presented here, since it is repulsive, is the
Ericson-Ericson-Lorentz-Lorenz (EELL) double scattering correction. 
However, we estimate that this effect, even though it may play an important role for the 
light isoscalar $\eta$ meson~\cite{Bass:2005hn}, is much reduced in the present case. This is 
because in the QMC model we work at the Hartree level and ignore the effect of 
nucleon correlations, or nonlocal interactions. Furthermore the EELL effect was aimed at 
low energy pion scattering with the assumption $qr\to 0$. 
For heavy mesons like the $\eta_c$ the momenta are much higher and the inverse
correlation length $<1/r>$ that appears in the EELL effect will certainly be much reduced by 
cancellations 
associated with the oscillatory behavior of the exponential. This plus the fact that we 
only work at the Hartree level and ignore exchange corrections that appear in 
the Hartree-Fock-based treatment.

\section{\label{sec:summary} Summary and discussion}

We have calculated the spectra of $\etac$-nucleus bound states for
various finite nuclei.  The meson-nucleus potentials were calculated using
a local density approximation, with the inclusion of the $DD^{*}$
meson loop in the $\etac$ self-energy.
The nuclear density distributions, as well as the in-medium $D$ and 
$D^{*}$ meson masses were consistently calculated by employing the
quark-meson coupling model. 
Using the meson potentials in nuclei, we have solved the Klein-Gordon 
equation and obtained meson--nucleus bound state energies. 
The sensitivity of our results to the 
cutoff parameter $\Lambda_{D}$ used in the vertex form factors
appearing in the $\etac$ self-energy has also been explored. 
Interestingly, the $\eta_c$-nucleus bound state energies calculated here are larger
than the corresponding $J/\Psi$ energies calculated in 
Ref.~\cite{Tsushima:2011kh}, by some of us, using the same approach. Needless to say, this deserves further investigation. 

Our results show that one should expect the $\etac$ to form bound states for all the nuclei studied, even though  the precise values of the bound state energies are dependent on the  cutoff mass values used in the form factors. The discovery of such bound states  would represent an important step forward in our understanding of the  nature of strongly interacting systems.

\section*{Acknowledgements}
This work was partially supported by Conselho Nacional de Desenvolvimento 
Cient\'{i}fico e Tecnol\'{o}gico (CNPq), process Nos.~313063/2018-4 (KT), 
426150/2018-0 (KT) and 309262/2019-4 (GK), and Funda\c{c}\~{a}o 
de Amparo \`{a} Pesquisa do Estado de S\~{a}o Paulo (FAPESP) process 
Nos.~2019/00763-0 (KT), 64898/2014-5 (KT) and 2013/01907-0 (GK). The work 
is also part of the project Instituto Nacional de Ci\^{e}ncia e Tecnologia 
-- Nuclear Physics and Applications (INCT-FNA), process. No.~464898/2014-5 
(KT, GK). It was also supported by the Australian Research Council through 
DP180100497 (AWT).



\end{document}